\documentclass[12pt]{article}
\usepackage{psfrag,epsfig}
\usepackage{a4,isolatin1,amsmath}
\begin{document}
\large \begin{center}
{\bf Instantaneous Spreading and Einstein Causality in Quantum Theory}
           \footnote{Talk given at the workshop {\em 
Superluminal\,(?)\,Velocities}, Cologne, June 6-10, 1998. A talk with a
             similar content was also given by the 
             author at the symposium {\em Uncertain Reality,}
             New Delhi, January 5-9, 1998.}
\end{center} \normalsize

\vspace*{0.3cm}

\begin{center}
{\bf Gerhard C. Hegerfeldt}\\
Institut f\"ur Theoretische Physik\\
Universit\"at G\"ottingen\\
Bunsenstr. 9\\
37073 G\"ottingen, Germany
\end{center}


\begin{abstract}
In nonrelativistic
quantum mechanics the wave-function of a free particle which initially
is in a finite volume immediately spreads to infinity. In a
nonrelativistic theory this is of no concern, but we show that the
same instantaneous spreading can occur in relativistic quantum theory
and that transition probabilities in widely separated systems may
instantaneously become nonzero. We discuss how this affects Einstein
causality.
\end{abstract}
\noindent \section{Introduction}
The principle of the
limiting role of the velocity of light is often called `Einstein
causality'\cite{tachyon}.
Einstein's principle of finite signal velocity is of
fundamental importance for the foundations of physics, both in
classical as well as in quantum physics.

Now consider nonrelativistic quantum mechanics and a free particle
whose wave function at time $t = 0$ vanishes outside some finite
region $V$. The latter
implies that the particle is in  $V$ with
probability 1, and thus with certainty. The free nonrelativistic
time evolution is then such that at an arbitrarily short  time later the
wave function is nonzero arbitrarily far away from the original region
$V$ \cite{Simon}. Thus the wave function  instantaneously spreads to
infinity and the probability of finding the particle arbitrarily
far away from the initial region is nonzero for any $t > 0$. But since
this superluminal propagation happens in a nonrelativistic theory it is of
no great concern.

Relativistic wave equations which are used to describe free
relativistic particles are hyperbolic and therefore the propagation is
at most luminal so that the above problem does not seem to
arise. However, if a wave function $\varphi ({\bf x}, t)$ satisfies a
relativistic wave equation then the connection between the variable
{\bf x} and the position of the particle is in general not clear-cut. 
In fact, the position operator proposed by Newton and Wigner 
\cite{Newton} is not multiplication by ${\bf x}$ but rather is a nonlocal
function of ${\bf x}$. It was noted by Weidlich and Mitra
\cite{Weidlich} and by Fleming \cite{Fleming} that
with the Newton-Wigner position operator superluminal propagation can occur in
the localization of a free relativistic particle. Amrein \cite{Amrein}
pointed out that this happened also for a proposed position operator
for the photon.  A similar
phenomenon occurred in some models in which localization was expressed
by means of a current-density four-vector as shown by Gerlach, Gromes,
and Petzold \cite{Petzold}.

In 1974 the present author \cite{He74} showed that this phenomenon of
instantaneous spreading  is quite general for a free relativistic
particle, 
irrespective of the particular notion of localization, be it in the
sense of Newton-Wigner or others. Later an alternative proof of this
result was given by Skagerstam \cite{Skag}, and it was extended to relativistic
systems by Perez and Wilde \cite{Perez} and to quite general, not
necessarily relativistic, interactions by Ruijsenaars and the author 
\cite{He80}. The upshot of Ref. \cite{He80} was that this instantaneous
spreading is mainly due to positivity of the energy plus translation 
invariance. Neither
relativity nor field theory is 
needed. It was recently shown by the  author \cite{Bohm} that
translation invariance is also not needed.  Hilbert space 
and positivity of the Hamiltonian (energy) suffices to ensure either
instantaneous spreading or, alternatively, confinement in a fixed
region for all times.

A further extension was given by the author \cite{He85} for free relativistic particles and for
relativistic systems which have exponentially bounded tails in their
localization outside some region $V$. It was shown that
the state spreads out to infinity faster than allowed by a probability
flow with finite propagation speed. 

In the following we give a brief   proof of our basic result of
Ref. \cite{He74}; for purely pedagogical reasons we use
relativistic invariance there. We then present an analysis of 
Fermi's two-atom system \cite{Fermi} 
where  we follow our Refs. \cite{He94,He95}. A large
portion will be devoted to a discussion what these results mean for
Einstein causality. Probably the most 
astonishing fact about our results is
that so little is needed to derive them. No field theory, no
relativity,  only  Hilbert space and
positivity of the energy is needed.

\section{ Spreading for particles}

If $\psi_t({\bf x})$ denotes a wave function of a particle in {\em
  non}relativistic quantum mechanics then the probability,
  $P_{\psi_t}(V)$, of finding the particle in a region $V$ at time $t$
 is given by
\begin{equation}\label{1}
P_{\psi_t}(V) = \int_V |\psi_t({\bf x})|^2 d^3 x \equiv \langle \psi_t,
\chi_V \psi_t \rangle
\end{equation}
where $\chi_V({\bf x})$ is the characteristic function of $V$ and
equals  $1$ in $V$ and $0$ outside. Regarded as
a multiplication operator, $\chi_V$ is a projector of the spectral
decomposition of the usual nonrelativistic position operator ${\bf
  X}$. For the Newton-Wigner position operators there are analogous
projection operators which determine $P_{\psi_t}(V)$, but which are no
longer given by a simple multiplication by a function.

In this section {\em no} position operator will be assumed nor any
form of projection operators. 
But rather I will assume just the {\em existence} of
some operator $N(V)$ such that the probability $P_{\psi_t}(V)$ of finding the
particle in $V$ at time $t$ is given by
\begin{equation}\label{2}
P_{\psi_t}(V) = \langle \psi_t, N(V)\psi_t \rangle
\end{equation}
where now $\psi_t$ is the (abstract) state vector of the free
relativistic particle. Because probabilities lie between $0$ and $1$ 
one must have 
\begin{equation}\label{3}
0 \le N(V) \le 1~.
\end{equation}
This implies in particular that $N(V)$ is hermitian and that
$N(V)^{1/2}$ exists as a hermitian positive operator. No explicit form
of $N(V)$ will be assumed in the following.

The time development of the free particle is described by the
generator $P^0 = \sqrt{{\bf P} + m^2}$ for time translation in an irreducible representation
of the Poincar\'e group for arbitrary mass $m \ge 0$ and spin $s =
0,~\frac{1}{2},...$~. By 
\begin{equation}\label{4}
U({\bf a}) = e^{- i {\bf P} \cdot {\bf a}/\hbar}
\end{equation}
we denote the operator for spatial translations. By translation
covariance, the probability of finding the translated state $U({\bf
  a}) \psi_t$ in the translated region $V_{\bf a}$ is the same as
that of finding $\psi_t$ in $V$. Replacing $\psi_t$ by $\psi_t^\prime
\equiv U(-{\bf a}) \psi_t$ the last statement implies
\begin{equation}\label{5}
P_{\psi_t}(V_{\bf a}) = \langle \psi_t, U({\bf a}) N(V) U({\bf
  a})^\dagger \psi_t \rangle~.
\end{equation}
Note that the r.h.s. of this can be written as a vector
norm. We can now prove the following results on instantaneous
spreading.\\

{\bf Theorem 1} \cite{He74}. Consider a free relativistic particle of
positive or zero mass and arbitrary spin. Assume that at time $t = 0$
the particle is localized with probability 1 in a bounded region $V$. 
Then there is a nonzero probability of finding the particle
arbitrarily far away at any later time.\\

{\bf Proof:} The proof is by contradiction. Let $t > 0$ be arbitrary
but fixed and
assume that at time $t$ 
the particle is localized with probability 1 in some
large, but finite, region containing $V$. This implies that for all 
sufficiently 
large $|{\bf a}|$, larger than some constant $R_t$ say, one has 
$P_{\psi_t}(V_{\bf a}) = 0$, as indicated in Fig. 1. 
\begin{figure}[h]
\begin{center}
\psfrag{V}{$V$}
\psfrag{Va}{$V_{\bf a}$}
\psfrag{ct}{$ct$}
\psfrag{a}{${\bf a}$}
\epsfig{file=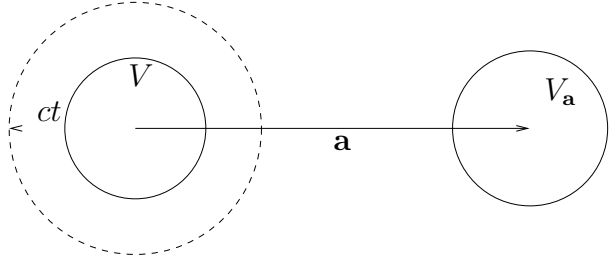,width=8cm}
\caption{ Maximal localization region at time $t$  has no
  intersection with $V_{\bf a}$.  }
\end{center}
\end{figure}
Eq.~(5) then becomes
\begin{equation}\label{6}
\|N(V)^{1/2} U({\bf a})^\dagger \psi_t \|^2 = 0~~{\rm for~all}~~ |{\bf
  a}| > R_t~.
\end{equation}
Thus the vector inside the norm vanishes. Multiplying the 
vector by $N(V)^{1/2}$ one obtains 
\begin{equation}\label{7}
N(V) U({\bf a})^\dagger \psi_t = 0~~{\rm for~all}~~ |{\bf a}| > R_t~.
\end{equation}
Taking the scalar product of this with $\psi_0$ yields
\begin{equation}\label{8}
\langle N(V) \psi_0, U(-{\bf a}) \psi_t \rangle = 0~~{\rm for}~~ |{\bf
  a}| > R_t~.
\end{equation}
We now introduce the abbreviations
\begin{eqnarray}\label{9}
\phi & \equiv & N(V) \psi_0\nonumber\\
f_t({\bf x}) & \equiv & \langle \phi, U(-{\bf x}) \psi_t \rangle~.
\end{eqnarray}
In momentum space ${\bf P}$ becomes multiplication by ${\bf p}$ and
thus the r.h.s. of Eq. (\ref{9}) can be written as
\begin{equation}\label{10}
f_t({\bf x}) = \int \underbrace{\frac{d ^3p}{(p^2 + m^2)^{1/2}}
    \boldsymbol{\phi} ({\bf p})^\star \cdot\boldsymbol{\psi}_0 ({\bf p})e^{- i
\sqrt{{\bf p}^2 + m^2}t}} e^{i {\bf p} \cdot {\bf x}}
\end{equation}
where a summation over spin variables is understood and where $m = 0$
is allowed.

Since, by Eq. (\ref{7}), $f_t(\boldsymbol{x})$ vanishes outside the ball of
radius $R_t$ around the origin (which means that $f_t$ has compact
support) its Fourier transform must be analytic, both for $t = 0$ and
the fixed $t > 0$ under consideration. But it is evident from
Eq. (\ref{10}) that the Fourier transform of $f_t(\boldsymbol{x})$ is the
expression over the brace. However, because of the root in the
exponent this expression cannot be analytic in {\bf p} for two distinct values of
$t$ unless $f_t \equiv 0$. For $t = 0$ and $\boldsymbol{x} = 0$ the latter
would imply
\[
\langle N(V) \psi_0, \psi_0 \rangle = 0,
\]
which is a contradiction. This concludes the proof.

It is worthwhile to point out that Theorem 1 carries over to unbounded
regions $V$ like infinite slabs of the form $\{(\boldsymbol{x}); |x_1| \le
d\}$ and that the spreading is indeed over all space (`no holes') and
relativity is not needed,
as shown by the author and Ruijsenaars \cite{He80}. Recently the
author\cite{Bohm} showed the stronger result 
that  Hilbert space and
positivity of the energy (Hamiltonian) yield instantaneous spreading
for systems.  

In Ref. \cite{He85} the author proved a stronger result for exponential tails.
We say that a particle is localized with exponentially bounded
tails if the probability of finding it outside a ball $B_r$ of radius
$r$ around the origin, i.e. in $I\!\!R^3 \backslash B_r$, decreases at
least as $\exp\{- Kr\}$  for large $r$, with some $ K > m \ge
0$. It was shown that if a particle is thus localized
at time $t = 0$
then at any later time $t$ one has, for any $c > 0$ and any $r_0 > 0,$
\begin{equation}\label{11}
P_{\psi_t}\left(I\!\!R^3 \backslash B_{r_0 + c t}\right) > P_{\psi_0} 
\left( I\!\!R^3 \backslash B_{r_0} \right)~.
\end{equation}
The strict inequality sign is the main point here. 
For relativistic systems the same result holds if the tails are
 bounded Gaussian-like with $\exp\{- K' r^2\}$, some $K' > 0$. 
\begin{figure}[h]
\begin{center}
\psfrag{r0}{$r_0$}
\psfrag{ct}{$ct$}
\epsfig{file=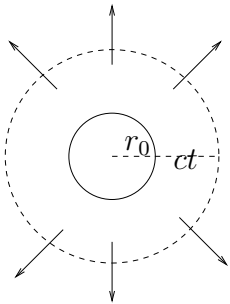,width=3cm}
\caption{Probability flow of speed at most $c$.}
\end{center}
\end{figure}

As seen from Fig. 2, if the change of localization probability with time were due to a
probability flow of speed at most $c$, then the probability of finding
the particle at time $t$ outside the ball $B_{r_0 + c t}$ could only
have grown by the amount between the two spheres and could therefore not
exceed that at time $t = 0$ outside the ball $B_{r_0}$. Just this is
violated in Eq. (\ref{11}). 

Eq. (\ref{11}) yields Theorem 1 if one chooses $V \subset B_{r_0}$ and
assumes $P_{\psi_0}(I\!\!R^3 \backslash V) = 0$. Then the r.h.s. of
Eq. (\ref{11}) is zero while the l.h.s. is nonzero so that the
probability of finding the particle outside any large ball in
nonzero. 

\section{The two-atom model of Fermi}

To check finite propagation speed Fermi \cite{Fermi} supposed in his
model that by some means one had prepared, at time $t = 0$, an atom
$A$ in an excited state $|e_A \rangle$ and some distance $R$ away an
atom $B$ in its ground state, with no photons initially present. The
state of the system then develops in time and Fermi calculated the
probability to find the state $|g_A \rangle |g_B \rangle |0_{{\rm ph}}
\rangle$ at time $t$. He probably had in mind that this only could
occur by de-excitation of $A$ accompanied by an emission of a photon
and then absorption of the latter by $B$ together with an excitation
of $B$. However, actually to check that there are no photons requires,
at least in principle, photon measurements over all space. Hence this
`exchange' probability cannot be used for signals, it just refers to
statistical correlations. In the spirit of Fermi a better approach
would be to calculate the probability of finding $B$ excited,
irrespective of the state of $A$ and possible photons. This approach
was analyzed by the present author in Refs. \cite{He94,He95}.

If the observable `$B$ is in an excited state' makes sense and is
represented by an operator ${\cal O}_{e_B}$ then 
\begin{equation}\label{12}
\langle \psi_t, {\cal O}_{e_B} \psi_t \rangle
\end{equation}
is the probability to find $B$ excited. Again, since probabilities lie
between zero and 1, one must have
\begin{equation}\label{13}
0 \le {\cal O}_{e_B} \le 1~.
\end{equation}
For example, ${\cal O}_{e_B}$ might be a projection operator, as for bare
states, but more general possibilities are conceivable. The explicit
form of ${\cal O}_{e_b}$ is not needed, only Eqs. (\ref{12}) and
(\ref{13}) are used. 

To more clearly separate what is physics and what mathematics the next
theorem is phrased in purely mathematical terms. Application to
physics will be explained subsequently.\\

{\bf Theorem 2} \cite{He94}. Let the operator $H$ be self-adjoint and
bounded from below. Let ${\cal O}$ be any operator satisfying
\begin{equation}\label{14}
0 \le {\cal O} \le {\rm const.}
\end{equation}
Let $\psi_0$ be any vector and define
\[
\psi_t \equiv e^{- i H t} \psi_0~.
\]
Then one of the following two alternatives holds.
\begin{itemize}
\item[(i)] $\langle \psi_t, {\cal O} \psi_t \rangle \not= 0 \qquad \qquad$
  for almost all $t$\\
(and the set of such $t$'s is dense and open)
\item[(ii)] $\langle \psi_t, O \psi_t \rangle \equiv 0$ \qquad \qquad
  for all $t$~.
\end{itemize}

The proof can be found in Refs. \cite{He94,He95}. In the application
to Fermi's model $H$ would be the Hamiltonian, ${\cal O} = {\cal O}_{e_B}$ and
$\psi_0$
 would represent the initial physical state in which $A$ is excited,
 $B$ not, and with no photons so that 
 $\langle \psi_0, {\cal O}_{e_B} \psi_0 \rangle =0$. 
One would then conclude that $B$ is
 either immediately excited with nonzero probability, or never.

The above mathematical result  has been used recently by the
author \cite{Bohm} to show that translation invariance is not needed
for instantaneous spreading of particles or system. The Hamiltonian
can be quite general, only boundedness from below is required, and this
ensures either instantaneous spreading or confinement in a fixed bounded
region for all times, corresponding to the alternatives (i) and (ii) above.

\section{Spreading in relativistic quantum mechanics}

At first sight the Dirac equation might seem to be a counterexample to
our results on instantaneous spreading. Indeed, this wave equation is
hyperbolic, implying 
finite propagation speed. For the localization operator $N(V)$ 
one might take  the characteristic function  $\chi_V(\boldsymbol{x})$,
just as in the nonrelativistic case of Eq. (\ref{1}) and in contrast to the
(nonlocal) Newton-Wigner operator. Then, for a wave function initially
vanishing  outside a finite region, i.e. of finite support, the
localization does evolve with finite propagation speed! Doesn't this
contradict Theorem 1 above?

This example is instructive since it shows the importance of the
positive-energy condition. The Dirac equation contains positive and
negative energy states, and therefore we conclude from our results
that positive-energy solutions of the Dirac equation always have {\em
  infinite} support to begin with! This is phrased as a mathematical
result for instance in the book of Thaller \cite{Thaller}. 

If there are no strictly localized states in the theory then Theorem 1
does not apply! Strict localization of a state $\psi$ in a region $V$
means that $\langle \psi, N(V) \psi \rangle = 1$, and this gives
\[
0 = \langle \psi, ({\bf 1} - N(V)) \psi \rangle = \| ({\bf 1} -
N(V))^{1/2} \psi \|^2
\]
where the root exists by Eq. (\ref{3}). Similar to Eqs. (\ref{6}) and
(\ref{7}) this implies
\begin{equation}\label{14a}
N(V) \psi = \psi .
\end{equation}
Hence $\psi$ is an eigenvector of $N(V)$ for the eigenvalue 1 if
$\psi$ is strictly localized in $V$, and vice versa. The eigenvalue
$0$ means strict localization outside $V$.

The existence or nonexistence of strictly localized states depends on
the form of $N(V)$. For example, if one has a self-adjoint position
operator $\hat{\boldsymbol{X}}$ with commuting components, then $N(V)$ is a
projection operator from the spectral decomposition of
$\hat{\boldsymbol{X}}$
and thus has eigenvalues 1 and $0$. Hence in this case there are
strictly localized states for any region $V$, and Theorem 1 implies
instantaneous spreading.

This instantaneous spreading also occurs for position operators with
self-adjoint but {\em non-}commuting components $\hat{X}_i$. Each
$\hat{X}_i$ has a spectral decomposition whose projection operators
give the localization operators for infinite slabs. Eigenvectors for
the eigenvalue 1 represent states strictly localized in these slabs,
and by the remark after Theorem 1 there is instantaneous
spreading in this case, too.

To avoid instantaneous spreading one has therefore to consider
localization operators $N(V)$ which are not projectors, for example
positive operator-valued measures. However, if one insists on
arbitrary good localization, i.e. on tails which drop off arbitrarily
fast, then one runs into Eq. (\ref{11}). This equation can be
interpreted as stemming from an infinitely fast probability flow. 

If in the Fermi model atoms $A$ and $B$ would develop instantaneous tails
then an immediate excitation would not seem surprising. So the two
phenomena appear to be connected. 
Mathematically this connection is of course
given by the assumed positivity of the energy and has been discussed
in detail by the author in Ref. \cite{Bohm}.

Could instantaneous spreading be used for the
transmission of signals if it occurred in the framework of
relativistic one-particle quantum mechanics? 
Let us suppose that at time $t=0$ one could prepare an ensemble of
strictly localized 
(non-interacting) particles by laboratory means, e.g. photons in an
oven. Then one could open a window  and would  observe
some of them at time $t = \varepsilon$ later on the moon. Or to better
 proceed by repetition, suppose one could 
 successively prepare strictly
localized individual particles in the laboratory. Preferably this
should be done with 
different, distinguishable, particles in order to be sure when a
detected particle was originally released. Such a signaling procedure
would have very low efficiency but still could be used for
synchronization of clocks or, for instance, for betting purposes. 

\section{Field-theoretic aspects and discussion}

{\em Localization of particles}. In field theory difficulties with particle
localization have been known for a long time 
\cite{Schweber}. However, our results are more general since 
no fields are involved and none of the particular assumptions or 
axioms of field theory, except for positivity  of the energy, are used.

In a field-theoretic context permanent infinite tails can  
be understood intuitively through clouds of virtual particles due to
renormalization (`dressed states'). It is also conceivable that
whenever one tries to prepare a localized particle or system one
automatically creates particle-antiparticle pairs outside the desired
localization region, and this would have the same effect as tails.

Instead of speaking about infinite tails one may also envisage that
all particle detectors exhibit, for example, inherent noise and that
therefore localization with probability 1 or zero can never be
recorded. This would essentially lead to the same conclusions as
permanent infinite tails.

{\em Fermi's model}.  From the foregoing discussion it is
evident that, 
in a field-theoretic context, excitation of atom $B$ need not be due
to absorption of a photon emitted by atom $A$. The excitation could
rather be due to vacuum fluctuations, photon clouds etc. Or the
excitation maybe just
spontaneous, whatever that means. At a more fundamental level, the
notion of bounds states has its intricacies in field theory, and the
corresponding observables in the Fermi model might be hard to put on a
mathematically rigorous footing.

If infinite tails always exist, or if all counters are influenced by
vacuum fluctuation, then how can finite propagation speed or Einstein
causality be checked at all? Here it is useful for differentiate
between two notions of causality.

{\em Strong causality}. By this we mean that for {\em each}
individual experiment in which two systems, separated by a distance
$R$, are prepared at time $t = 0$ {\em no} disturbance or excitation of the
second system occurs for $t < R/c$ \cite{He95}. This notion is
analogous to energy 
momentum and baryon conservation in each individual scattering process
in particle physics. Strong causality would hold in the Fermi model if
the transition probability were strictly zero for $t < R/c$. It seems
that Fermi had this causality notion in mind. Our results show that
strong causality cannot be {\em checked} (unless a possible way out
via cut-off theories holds \cite{He95}), or it may fail in a
theory.

{\em Weak causality}. This notion was introduced by Schlieder
\cite{Schlieder}, and it means Einstein causality for expectation
values or ensemble averages only, not for each individual
process. Thus for the above two systems, expectation values for the
second system need not vanish for $t < R/c$, but it would take at least a
time $t = R/c$ to produce an effect on them. To exhibit this effect
it is convenient to subtract possible fluctuations of the second
system alone. Theorem 2 does not apply to this situation since this
difference is not the expectation value of a positive operator.

The weak assumptions of the above theorems (just Hilbert space and positive
energy) will not be enough to prove weak causality. In
Refs. \cite{LP,MJF} the Fermi model has been studied  using the methods
and approximations of quantum optics. Vacuum fluctuations and virtual
photons contribute to the excitation of atom $B$, and once the
expectation value of this contribution has been subtracted, the
remainder behaves causally, at least within the approximations
employed. This just corresponds to the notion of weak causality. In
how far this can be measured will be discussed below.

In the framework of quantum field theory it is sometimes simply argued
that local commutation or anti-commutation relations must clearly
ensure causal behavior, for instance of localized particles. This
implicitly presupposes, however, that the relevant operators,
e.g. $N(V)$ or ${\cal O}_{e_B}$ --if they exist -- are local functions of
the fields. It should be recalled that fields do not enter in our
formulation, and it should be noted that if $N(V)$ is a local
function of the fields then the Reeh-Schlieder theorem \cite{Reeh}
implies that its vacuum expectation is nonzero.

Within local quantum field theory a rigorous proof of {\em weak}
causality for {\em local} observables has been given by Schlieder
\cite{Schlieder} and by Buchholz and Yngvason \cite{BY}, as well as by
Neumann and Werner \cite{Neumann} in an alternative algebraic
framework. In Ref. \cite{BY} it was moreover shown that 
 restrictions of states to local algebras cannot be tested by means of
 transition probabilities.

How can one check weak causality experimentally? In the Fermi model
one would not use a single pair of atoms but rather an ensemble,
either by repetition or by simultaneous realization. If in the Fermi
case one has $N$ pairs of atoms $A$ and $B$ one would measure at time
$t$ how many $B$ atoms are excited. For $N \rightarrow \infty$ their
fraction would be given by $\langle \psi_t, {\cal O}_{e_B} \psi_t
\rangle$, while for finite $N$ this would hold only approximately, due
to statistical fluctuations. Then one would subtract the -- either
calculated or measured -- excitation probability of $B$ without $A$
present. Weak causality asserts that the difference should be zero for
$t < R/c$ -- but only for $N \rightarrow \infty$, while for finite $N$
there are always fluctuations. Hence in a strict sense, weak Einstein
causality can only be checked experimentally for infinite
ensembles. This suggests a macroscopic context.\\

 The main point of our results seems to be that instantaneous
 spreading holds  already under amazingly few  assumptions.
Neither the existence of fields nor the usual
axioms of field theory are assumed -- the only input is Hilbert space
and positivity of the energy. Our results seem to indicate the need
for  a mechanism like  vacuum fluctuations,
clouds of virtual particles, particle-antiparticle pairs, spontaneous
excitations, or something like that in order to retain Einstein
causality.  Our results are compatible with quantum field theory which
uses much stronger assumptions and in which vacuum fluctuations etc.
 are present.

\end{document}